\documentclass[sigconf]{acmart}

\AtBeginDocument{%
  \providecommand\BibTeX{{%
    \normalfont B\kern-0.5em{\scshape i\kern-0.25em b}\kern-0.8em\TeX}}}

\acmConference[Conference '19]{ACM}{December 2019}{USA}

\usepackage{amssymb}
\usepackage{amsmath}
\usepackage{amsfonts}
\usepackage[mathscr]{euscript}
\usepackage{multirow}
\usepackage{url}
\usepackage{subfig}
\usepackage{multicol}
\usepackage{float}
\usepackage{graphicx}
\usepackage{caption}

\usepackage{tikz}
\newcommand{\empt}[2]{$#1^{\langle #2 \rangle}$}
\usetikzlibrary{
  arrows.meta, 
  fit, 
  positioning, 
  shapes, %
  calc,
  chains,
}
\tikzset{
  neuron/.style={ 
    circle,draw,thick, 
    inner sep=0pt, 
    minimum size=3.5em, 
    node distance=1ex and 2em, 
  },
  group/.style={ 
    rectangle,draw,thick, 
    inner sep=0pt, 
  },
  io/.style={ 
    neuron, 
    fill=gray!15, 
  },
  conn/.style={ 
    -{Straight Barb[angle=60:2pt 3]}, 
    thick, 
  },
}
\tikzset{elementwiseoperation/.style={circle, draw, inner sep=0pt},
    elementwisefunction/.style={ellipse, draw, inner sep=1pt},
    ct/.style={circle, draw, minimum width=1cm, inner sep=1pt},
    gt/.style={rectangle, draw, minimum width=4mm, minimum height=3mm, inner sep=1pt},
    mylabel/.style={font=\scriptsize\sffamily},}

\usepackage{soul}
\usepackage{epsfig}
\newcommand{\maian}{\textsc{Maian}}
\usepackage{enumitem}

\begin{document}

\title[Safer Smart Contracts: A Sequence Learning Approach]{Towards Safer Smart Contracts: A Sequence Learning Approach to Detecting Security Threats}

\author{Wesley Joon-Wie Tann}\affiliation{%
  \institution{National University of Singapore}
}
\email{wesleyjtann@u.nus.edu}
\authornote{The first author performed a large part of the work while at Nanyang Technological University}

\author{Xing Jie Han}
\affiliation{%
  \institution{Nanyang Technological University}
}
\email{xhan017@e.ntu.edu.sg}

\author{Sourav Sen Gupta}
\affiliation{%
  \institution{Nanyang Technological University}
}
\email{sg.sourav@ntu.edu.sg}

\author{Yew-Soon Ong}
\affiliation{%
  \institution{Nanyang Technological University}
}
\email{asysong@ntu.edu.sg}


\begin{abstract}
%
Symbolic analysis of security exploits in smart contracts has demonstrated to be valuable for analyzing predefined vulnerability properties. While some symbolic tools perform complex analysis steps, they require a predetermined invocation depth to search vulnerable execution paths, and the search time increases with depth. The number of contracts on blockchains like Ethereum has increased 176 fold since December 2015. If these symbolic tools fail to analyze the increasingly large number of contracts in time, entire classes of exploits could cause irrevocable damage. In this paper, we aim to have safer smart contracts against emerging threats. We propose the approach of sequential learning of smart contract weaknesses using machine learning---long-short term memory (LSTM)---that allows us to be able to detect new attack trends relatively quickly, leading to safer smart contracts. Our experimental studies on 620,000 smart contracts prove that our model can easily scale to analyze a massive amount of contracts; that is, the LSTM maintains near constant analysis time as contracts increase in complexity. In addition, our approach achieves $99\%$ test accuracy and correctly analyzes contracts that were false positive (FP) errors made by a symbolic tool.


\end{abstract}



\maketitle

\section{INTRODUCTION}
Smart contracts provide automated peer-to-peer transactions while leveraging on the benefits of the decentralization provided by blockchains. As smart contracts are able to hold virtual coins worth upwards of hundreds of USD each, they have allowed the automated transfer of monetary values or assets via the logic of the contract while having the correctness of its execution governed by the consensus protocol~\cite{Nakamoto_bitcoin}. The inclusion of automation in blockchain resulted in rapid adaptation of the technology in various sectors such as finance, healthcare, and insurance~\cite{zile2018blockchain}. Ethereum, the most popular platform for smart contracts had a market capitalization upwards of $\$21$ billion USD~\cite{ethereumprice}. Due to the fully autonomous nature of smart contracts, exploits are especially damaging as they are largely irreversible due to the immutability of blockchains. On Ethereum alone, over 3.6 million Ether (virtual coins used by Ethereum) were stolen from a decentralized investment fund called The DAO (Decentralized Autonomous Organization) in June 2016, incurring losses of up to $\$70$ million USD~\cite{TheDAO}. In November 2017, $\$300$ million USD was frozen because of Parity's MultiSig wallet~\cite{Parity}. Both hacks that occurred were due to exploitable logic within smart contracts themselves, and these incidents highlighted a strong imperative for the security of smart contracts.

The tools used in smart contract symbolic analysis are mainly based on formal methods of verification. While most analysis tools have applied dynamic analysis to automatically detect bugs in smart contracts~\cite{Luu:2016:MSC:2976749.2978309, DBLP:conf/ndss/KalraGDS18, Tsankov:2018:SPS:3243734.3243780}, some have focused on finding vulnerabilities across multiple invocations of a contract~\cite{Liu:2018:RFR:3183440.3183495}. \textsc{Oyente} is one such example of an automatic bug detector. It was proposed to act as a form of pre-deployment mitigation, by analyzing smart contracts for vulnerabilities at a bytecode level~\cite{Luu:2016:MSC:2976749.2978309}. It uses symbolic execution to capture traces that match the characteristics of the classes of vulnerabilities as defined. However, it is not complete as confirmations of flagged contracts being vulnerable were only done manually in the presence of contract source code. 

Recently, it has been shown that \maian, a tool for precisely specifying and reasoning about trace properties, which employs inter-procedural symbolic analysis and concrete validation for exhibiting real exploits~\cite{Nikolic:2018:FGP:3274694.3274743}, was able to capture many well-known examples of unreliable bugs. Using predefined execution trace vulnerabilities directly from the bytecode of Ethereum smart contracts, \maian\ labels vulnerable contracts as one or two of the three categories---suicidal, prodigal, and greedy. \maian\ is able to detect different classes of vulnerabilities that may only appear after multiple invocations while verifying its results on a private fork of Ethereum. However, the degree of accuracy in its detection is limited by its invocation depth, whereby states that vulnerabilities may occur in were not reached due to a tradeoff between analysis time and exhaustiveness of search. In addition, concrete validation of contracts can only be performed by \maian\ either on flagged contracts that are alive within the forked Ethereum chain or on contracts with existing source code readily available. 

In the field of machine learning, recurrent neural networks are exceptionally expressive and powerful models adapted to sequential data. The long short-term memory (LSTM) model is a compelling variant of recurrent networks mainly used to solve difficult sequential problems such as speech recognition~\cite{hinton2012deep, Dahl:2012:CPD:2335874.2336015}, machine translation~\cite{cho-al-emnlp14, DBLP:journals/corr/WuSCLNMKCGMKSJL16}, and natural language processing~\cite{journals/corr/Graves13, luong-etal-2015-effective}. In recent years, there has been an increasing interest in the security of smart contracts and the application of machine learning in computer security, with papers on automated exploit analysis~\cite{217650}, neural networks for guessing passwords~\cite{204155}, exploits for a contract developed from bytecode~\cite{217464}, and taxonomy of common programming pitfalls~\cite{Atzei:2017:SAE:3080353.3080363}.

In this work, we introduce an LSTM model for detecting smart contract security threats at an opcode level. To the best of our knowledge, this is the first machine learning approach to smart contract exploit detection. We study the applicability of using an LSTM model to detect smart contract security threats. As smart contracts become available in sequential order, they could be used to update the LSTM model for future contracts at each point in time. Since only around $1\%$ of all smart contracts have available Solidity source code (we refer to Etherscan~\cite{etherscan}), it highlights the utility of our LSTM learning model as a smart contract security tool which operates solely at the opcode level. 

{\flushleft\textbf{Contributions}.} Our contributions in this work are as follows:
\begin{itemize}
	\item We show that our LSTM model is able to outperform symbolic analysis tool \maian~\cite{Nikolic:2018:FGP:3274694.3274743} in detecting smart contract security vulnerabilities. 
	\item We experimentally demonstrate that the LSTM performance improves with new contracts, and eventually achieved 
	\begin{enumerate}[leftmargin=2\parindent]
	    \item detection test accuracy of $99.57\%$
	    \item $F_1$ score of $86.04\%$
	\end{enumerate}
    \item We show that our approach detected up to $92.86\%$ of challenging contracts that were false positive (FP) errors made by \maian.
    \item We show that the LSTM, which only requires constant analysis time as smart contracts grow in complexity, can easily scale to process a large number of smart contracts.
    \item By demonstrating that the proposed LSTM tool is a competitive alternative to symbolic analysis tools, we set a benchmark for future work on machine learning models that ensure smart contracts security.
\end{itemize}

\section{Background}
\subsection{Smart Contracts} \label{sec:smartcontracts}
Smart contracts are autonomous, state-based executable code that is stored, verified, and executed on the blockchain. Ethereum smart contracts are predominantly written in a high-level programming language---Solidity, which is then compiled to a stack-based bytecode format. A smart contract is deployed on the Ethereum blockchain in the form of a transaction by a sender, where an address is assigned to the contract. Each smart contract contains a state (account balance and private storage) and executable code. Once deployed, a smart contract is immutable and no modifications can be made to the contract. However, it may be killed if a \texttt{Suicide} instruction in the contract is executed. 

Contracts, once deployed on the blockchain, may be invoked by sending transactions to the contract addresses, along with input data and \emph{gas} (``fuel'' for smart contract execution). In Ethereum, gas is assigned proportionately to the amount of computation required for each instruction in its instruction set~\cite{wood2014ethereum}. This gas is used as an incentive within the proof-of-work system for executing the contracts. If gas is insufficient or exhausted before the end of execution, no gas is refunded to the caller and the transaction (including state) is reverted. No transactions can be sent to or from a killed contract. 

In Ethereum, an invocation of a smart contract is executed by every fullnode in the network, taking into account both the current state of the blockchain and the state of the executing contract, to reach consensus on the output of the execution. The contract would then update the contract state, transfer values to other contract addresses, and/or execute functions of other contracts.

\subsection{Contracts with Vulnerabilities} \label{sec:vulnerablecontracts}
Due to the autonomy and immutability of smart contracts, once an attack is executed successfully on a contract, it is impossible for the transaction to be reversed without performing a hardfork~\cite{hard_fork} on the underlying blockchain. As the distribution of smart contracts within Ethereum is heavily skewed towards the financial sector (primarily used for the transfer of assets or funds)~\cite{DBLP:conf/fc/BartolettiP17a}, some of the past attacks have incurred multimillion-dollar losses. This highlights a strong need for security of smart contracts.  Although there are several existing studies and analyses of exploit categories in smart contracts~\cite{Luu:2016:MSC:2976749.2978309, DBLP:conf/ndss/KalraGDS18, SmartInspect}, we primarily focus on the classes defined in~\cite{Nikolic:2018:FGP:3274694.3274743}, due to their extensive coverage and availability of the open source tool \maian. We will briefly go over some of the concepts and the exploit categories highlighted in the paper. 

An execution trace of a smart contract is a series of contract invocations that occurred during its lifetime. Exploits that happen over a sequence of contract invocations are known as \emph{trace exploits}. In~\cite{Nikolic:2018:FGP:3274694.3274743}, the exploits in Ethereum smart contracts are classified under three categories---suicidal, prodigal, and greedy. 

\paragraph{Suicidal Contracts.} Smart contracts that can be killed by any arbitrary address are classified as \emph{suicidal}. Although some contracts have an option to kill themselves as mitigation against attacks, if improperly implemented, the same feature may allow any other user the option of killing the contract as well. This occurred during the ParitySig attack~\cite{Parity}, where an arbitrary user managed to gain ownership of a library contract and killed it, rendering any other contract that relied on this library useless and effectively locking their funds. 

\paragraph{Prodigal Contracts.} Smart contracts classified as \emph{prodigal} are ones that can leak funds to arbitrary addresses, which either (a) do not belong to the owner of the contract, or (b) have not deposited Ether to the contract. Contracts often have internal calls to send funds to other contracts or addresses. However, if there are insufficient mechanisms in place to guard the availability of such calls, attackers may be able to exploit this call to funnel Ether to their own accounts, draining the vulnerable contract of its funds.

\paragraph{Greedy Contracts.}\label{greedy} Smart contracts that are unable to release Ether are classified as \emph{greedy}. Following the ParitySig attack~\cite{Parity}, many accounts dependent on the library contract were unable to release funds, resulting in an estimated loss of \$30 million USD. Within the greedy class, the vulnerable contracts are subdivided into two categories---(a) contracts that accept Ether but completely lack instructions to send funds, and (b) contracts that accept Ether and contain instructions to send funds, but are unable to perform the task. 

\subsection{Recurrent Neural Networks}
Recurrent Neural Networks (RNNs) are powerful machine learning models adapted to sequence data. These models can learn and achieve outstanding performance on many hard sequential learning problems such as speech recognition, machine translation, and natural language processing. These neural networks possess a remarkable ability to learn highly accurate models using only two hidden layers~\cite{conf/ijcnn/NguyenW90}. However, standard RNNs are hard to properly train in practice. The main reason why the model is so unmanageable is that it suffers from both exploding and vanishing gradients~\cite{BengioSimardFrasconi94}. Both issues are due to the RNN's recurrent nature. 

While the exploding gradients problem is relatively easy to solve by simply shrinking gradients with norms passing a certain threshold, a method known as gradient clipping~\cite{pascanu2013difficulty, mikolov2012statistical}, the vanishing gradient issue is much more challenging. This is because vanishing gradients do not cause the gradient itself to be small. In fact, the gradient's component in directions that correspond to short-term dependencies is large, while the component in directions that correspond to long-term dependencies is small. As a result, recurrent networks are able to easily learn short-term dependencies but not the long-term ones. 

\subsection{Long Short-Term Memory} \label{sec:lstm_background}
In order to address the vanishing gradient and long-term dependency issues of standard RNNs, the long short-term memory (LSTM) network was proposed~\cite{Gers:99c, hochreiter1997long}. In the LSTM, gate functions were recommended to be used for controlling information flow in any given recurrent unit---an input gate, a forget gate, and an output gate. An input gate functions as a gatekeeper to allow relevant signals through into the hidden context. On the other hand, the forget gate is used to determine the amount of prior information remembered for the current time-step, and the output gate functions as a prediction mechanism. By introducing such information gate controls, the LSTM almost always performs much better than standard RNNs. 

RNNs take a sequence $\{x_1, x_2, \ldots, x_T\}$ as input and construct a corresponding sequence of hidden states (or representations) $\{h_1, h_2, \dots, h_T\}$. In the simplest case, a single-layer recurrent network uses the hidden representations $\{h_1, h_2, \dots, h_T\}$ for estimation and prediction. In deep RNNs, each hidden layer uses the hidden states of the previous layer as inputs. That is, the hidden states in layer $k-1$ are used as inputs to layer $k$. In RNNs, every hidden state in each layer performs memory-based learning to place importance on relevant features of the task using previous inputs. Previous hidden states and current inputs are transformed into a new hidden state, and it is achieved through a recurrent operator that takes in $(h_{t-1}, x_t)$, such as:
\begin{equation*}\label{eq:Roperator}
	h_t = \tanh(W_h h_{t-1} + W_x x_t + b),
\end{equation*}
where $W_h$, $W_x$, and $b$ are parameters of the layer and $\tanh(\cdot)$ represents the standard hyperbolic tangent function. 

\begin{figure}[htbp]
\begin{center}
\resizebox {.95\columnwidth} {!} {
\begin{tikzpicture}[
    font=\sf \scriptsize,
    >=LaTeX,
    cell/.style={
        rectangle, 
        rounded corners=5mm, 
        draw,
        very thick,
        },
    operator/.style={
        circle,
        draw,
        inner sep=-0.5pt,
        minimum height =.2cm,
        },
    function/.style={
        ellipse,
        draw,
        inner sep=1pt
        },
    ct/.style={
        circle,
        draw,
        line width = .75pt,
        minimum width=1cm,
        inner sep=1pt,
        },
    gt/.style={
        rectangle,
        draw,
        minimum width=4mm,
        minimum height=3mm,
        inner sep=1pt
        },
    mylabel/.style={
        font=\scriptsize\sffamily
        },
    ArrowC1/.style={
        rounded corners=.25cm,
        thick,
        },
    ArrowC2/.style={
        rounded corners=.5cm,
        thick,
        },
    ]

    \node [cell, minimum height =4cm, minimum width=6cm] at (0,0){} ;

    \node [gt] (ibox1) at (-2,-0.75) {$\sigma$};
    \node [gt] (ibox2) at (-1.5,-0.75) {$\sigma$};
    \node [gt, minimum width=1cm] (ibox3) at (-0.5,-0.75) {Tanh};
    \node [gt] (ibox4) at (0.5,-0.75) {$\sigma$};

    \node [operator] (mux1) at (-2,1.5) {$\times$};
    \node [operator] (add1) at (-0.5,1.5) {+};
    \node [operator] (mux2) at (-0.5,0) {$\times$};
    \node [operator] (mux3) at (1.5,0) {$\times$};
    \node [function] (func1) at (1.5,0.75) {Tanh};

    \node[ct, label={[mylabel]Cell}] (c) at (-4,1.5) {\empt{c}{t-1}};
    \node[ct, label={[mylabel]Hidden}] (h) at (-4,-1.5) {\empt{h}{t-1}};
    \node[ct, label={[mylabel]left:Input}] (x) at (-2.5,-3) {\empt{x}{t}};

    \node[ct, label={[mylabel]}] (c2) at (4,1.5) {\empt{c}{t}};
    \node[ct, label={[mylabel]}] (h2) at (4,-1.5) {\empt{h}{t}};
    \node[ct, label={[mylabel]}] (x2) at (2.5,3) {\empt{h}{t}};

    \draw [ArrowC1] (c) -- (mux1) -- (add1) -- (c2);

    \draw [ArrowC2] (h) -| (ibox4);
    \draw [ArrowC1] (h -| ibox1)++(-0.5,0) -| (ibox1); 
    \draw [ArrowC1] (h -| ibox2)++(-0.5,0) -| (ibox2);
    \draw [ArrowC1] (h -| ibox3)++(-0.5,0) -| (ibox3);
    \draw [ArrowC1] (x) -- (x |- h)-| (ibox3);

    \draw [->, ArrowC2] (ibox1) -- (mux1);
    \draw [->, ArrowC2] (ibox2) |- (mux2);
    \draw [->, ArrowC2] (ibox3) -- (mux2);
    \draw [->, ArrowC2] (ibox4) |- (mux3);
    \draw [->, ArrowC2] (mux2) -- (add1);
    \draw [->, ArrowC1] (add1 -| func1)++(-0.5,0) -| (func1);
    \draw [->, ArrowC2] (func1) -- (mux3);

    \draw [-, ArrowC2] (mux3) |- (h2);
    \draw (c2 -| x2) ++(0,-0.1) coordinate (i1);
    \draw [-, ArrowC2] (h2 -| x2)++(-0.5,0) -| (i1);
    \draw [-, ArrowC2] (i1)++(0,0.2) -- (x2);

\end{tikzpicture}
}
\end{center}
\caption{Schematic of a Long Short-Term Memory Cell.}
\label{fig:LSTM_cell}
\end{figure}
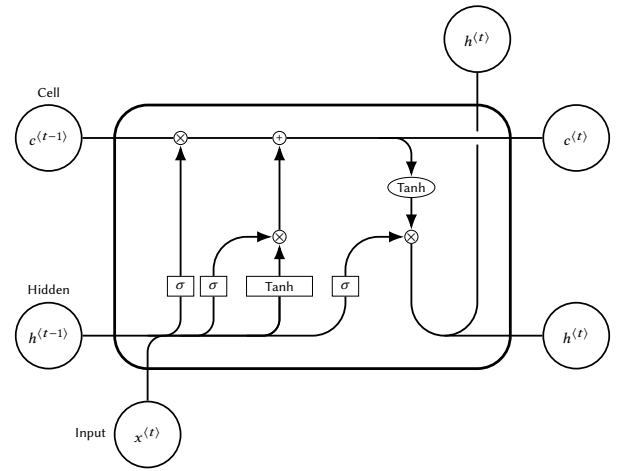

The LSTM architecture is specifically designed to handle recurrent operations. In this architecture, a memory cell $c_t$, as shown in Figure~\ref{fig:LSTM_cell}, is introduced for internal long-term storage. As we recall that the hidden state $h_t$ is an approximate representation of state at time-step $t$, both $c_t$ and $h_t$ are computed via three gate functions to retain both long and short term storage of information. The forget gate $f_t$, via an element-wise product, directly connects $c_t$ to the memory cell $c_{t-1}$ of the previous time-step. Using large values for the forget gates would cause the cell to retain almost all of its previous values. In addition, input gate $i_t$ and output gate $o_t$ control the flow of information within themselves. Each gate function has its own weight matrix and a bias vector. We denote the parameters with subscripts $f$ for the forget gate function, $i$ for the input gate function, and $o$ for the output gate function respectively (e.g., $W_{xf}$, $W_{hf}$, and $b_f$ are parameters of the forget gate function). 

Practitioners across various fields in sequence modeling use slightly different LSTM variants. In this work, we follow the model of leading natural language processing research~\cite{journals/corr/Graves13}, used to handle complex sequences with long-range structure. The following is the formal definition of our full LSTM architecture, without peep-hole connections,
\begin{align}
  &\begin{aligned}
    i_t &= \sigma(W_{xi}x_t + W_{hi}h_{t-1} + b_i)	
  \end{aligned}\\
  &\begin{aligned}
    f_t &= \sigma(W_{xf}x_t + W_{hf}h_{t-1} + b_f)
  \end{aligned}\\
  &\begin{aligned}
    o_t &= \sigma(W_{xo}x_t + W_{ho}h_{t-1} + b_o)	
  \end{aligned}\\
  &\begin{aligned}
    g_t &= \tanh(W_{xg}x_t + W_{hg}h_{t-1} + b_g)	
  \end{aligned}\\
  &\begin{aligned}
    c_t &= f_t \odot c_{t-1} + i_t \odot g_t	
  \end{aligned}\\
  &\begin{aligned}
    h_t &= o_t \odot \tanh(c_t)	
  \end{aligned}
\end{align}
where $\sigma(\cdot)$ is the sigmoid function, $\tanh(\cdot)$ is the hyperbolic tangent function, and $\odot$ denotes element-wise product.

\section{Learning Smart Contract Threats} \label{sec:learningsc}
In this section, we propose the modeling of smart contract exploits using a sequential machine learning approach, and explain how an LSTM learning model handles the semantic representations of smart contract opcode. We present the security threat detection objective, properties of smart contract opcode as a sequence, and opcode embedding representation. 

\subsection{Classification of Contract Threats}
The objective of our LSTM learning model is to perform a two-class classification, in order to detect if any given smart contract contains security threats. Motivated by the concepts in optimization, the objective in LSTM learning is to minimize the detection loss function, in order to maximize classification accuracy. Through the loss provided by each training data point, we ideally expect the sequence model to learn from the errors. Loss functions of learning models are mostly application specific and are selected based on how they affect the performance of the classifiers~\cite{Altun:2003:ILF:1119355.1119374,bishop1995neural}. The most common ones used to measure the performance of a classification model are the cross-entropy loss (logarithmic loss), softmax, and squared loss. 

\begin{figure}[htbp]
\begin{center}
\makebox[\linewidth]{\includegraphics[width=.95\linewidth]{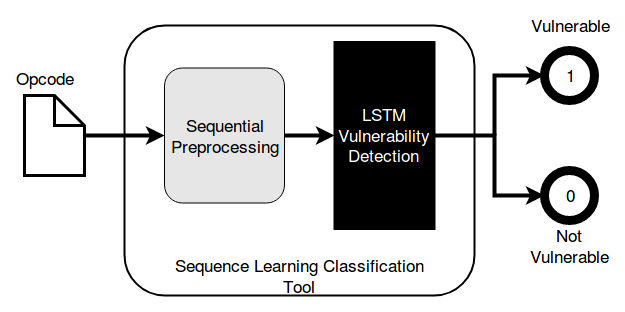}}
\end{center}
\caption{LSTM Smart Contract Vulnerability Classification.}
\label{fig:lstm_detect}
\end{figure}

In our case, we have chosen the logarithmic loss or the binary cross-entropy loss function. It is preferred as we formalized the smart contract threat detection into a binary classification problem. Let us proceed to define the derivation of a binary cross-entropy loss function $\mathscr{L}$:

\begin{equation}\label{eq:crossentropy}
	\mathscr{L} = -\frac{1}{N} \sum_{x}[y \ln a + (1-y) \ln (1-a)],
\end{equation}
where $N$ is the total number of contract opcodes in the training dataset, $x$ the sum over all training opcodes, $a = \sigma(z)$, where $z = \sum_{j} w_j x_j + x_{+1}$ is the weighted sum of the inputs, and $y$ the corresponding desired threat estimate. As the network improves its estimation of desired outputs $\mathbf{y}$ for all training opcodes $\mathbf{X}$, the summation of cross-entropy loss tends toward zero. This means that as a model learns to be more accurate in classifying smart contracts over time, it minimizes the distance between output estimate $a$ and the desired output $y$. A perfect classifier would achieve a log loss of precisely zero. 

\subsection{Sequential Modeling of Smart Contracts}
In this section, we first introduce the Ethereum opcode sequence processed by the LSTM model, followed by the usage of smart contract opcode sequence as input for our learning model to detect security threats, Figure~\ref{fig:lstm_detect}.

\subsubsection{Ethereum Opcode Sequence} \label{sec:EVMopcode}
Smart contract threat detection, like many sequence learning tasks, involves processing sequential opcode data. More precisely, opcodes are a sequence of numbers interpreted by the machine (virtual or silicon) that represents the type of operations to be executed. In the Ethereum environment, opcodes are a string of low-level human-readable instructions specified in the yellow paper~\cite{wood2014ethereum}. The machine instruction language is processed by Ethereum Virtual Machine (EVM)---a stack-based architecture with a word size of 256-bit. Each instruction is defined with an opcode (value), name (mnemonic), $\delta$ value, $\alpha$ value, and a description. For each instruction, the $\alpha$ value is the number of additional items placed on the stack for that instruction. Similarly, the $\delta$ value is the number of items required on the stack for that instruction.

\begin{figure}[ht]
\begin{center}
\resizebox{0.9\linewidth}{!}{\noindent\mbox{%
    \parbox{\linewidth}{\small 60 60 52 36 15 61 57 60 35 7c 90 04 63 16 80 63 14 61 57 80 63 14 61 57 80 63 14 61 57 5b 61 5b 60 60 90 54 90 61 0a 90 04 73 16 73 16 34 60 51 80 90 50 60 60 51 80 83 03 81 85 87 61 5a 03 f1 92 50 50 50 15 61 57 7f 60 60 90 54 90 61 0a 90 04 73 16 60 51 80 82 73 16 73 16 81 52 60 01 91 50 50 60 51 80 91 03 90 a1 61 56 5b 60 60 fd 5b 5b 56 5b 00 5b 34 15 61 57 fe 5b 61 60 80 80 35 73 16 90 60 01 90 91 90 50 50 61 56 5b 00 5b 34 15 61 57 fe 5b 61 60 80 80 35 73 16 90 60 01 90 91 90 50 50 61 56 5b 00 5b 61 60 80 80 35 73 16 90 60 01 90 91 90 80 35 90 60 01 90 82 01 80 35 90 60 01 90 80 80 60 01 60 80 91 04 02 60 01 60 51 90 81 01 60 52 80 93 92 91 90 81 81 52 60 01 83 83 80 82 84 37}%
}}
\end{center}
\caption{Sample opcode sequence used as input data to the LSTM learning model.}
\label{fig:opcode}
\end{figure}

\begin{figure}[hbp]
\begin{center}
\makebox[\linewidth]{\includegraphics[width=.95\linewidth]{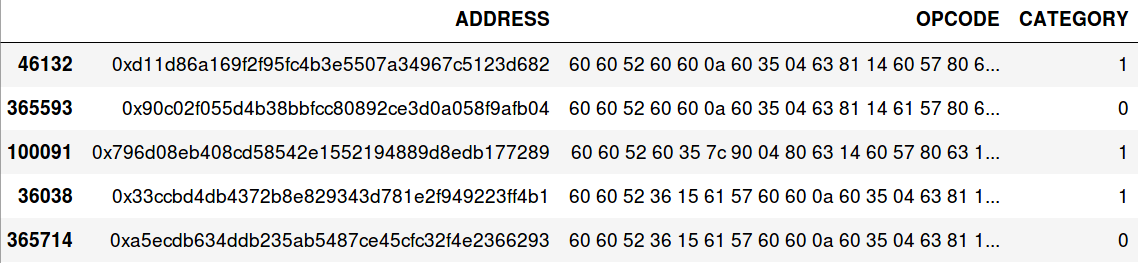}}
\end{center}
\caption{Dataset: Contract address, opcode, and category.}
\label{fig:data_head}
\end{figure}

To generate the labels required for supervised machine learning, the contracts were processed by passing bytecodes through \maian\ to obtain vulnerability classifications. In the process, opcodes were also retrieved. A sample EVM opcode thus produced, which the LSTM model takes as input is shown in Figure~\ref{fig:opcode}. The addresses of the contracts were saved, along with the valid corresponding EVM opcodes, and threat classifications (categories) into a data-frame, Figure~\ref{fig:data_head}.

We then use these smart contract opcodes as input to our sequence learning model. Our choice of using opcodes for learning smart contract threats is based on the long-proven capability of machine learning malware detection in both Windows and Android systems. In malware detection, models typically learn from opcode features to achieve impressive detection accuracy~\cite{Ngram_kang, Abou-Assaleh:2004:NDN:1025118.1025582}. This approach of learning from opcode features prevails over traditional malware detection approaches such as signature-based detection and heuristic-based detection, even offering the added benefit of being able to learn from existing patterns at a binary level to classify unknown threats~\cite{Shabtai:2009:DMC:1550969.1551289}. In this study, we propose a similar approach by applying machine learning to opcodes derived from Ethereum smart contracts.

\subsubsection{Opcode Sequence for Threat Detection} \label{sec:seq}
\begin{figure}[htp]
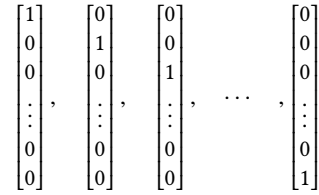

\begin{center}
    \[
\begin{bmatrix}
    1       \\
    0       \\
    0       \\
    \vdots  \\
    0       \\
    0
\end{bmatrix}, \quad
\begin{bmatrix}
    0       \\
    1       \\
    0       \\
    \vdots  \\
    0       \\
    0
\end{bmatrix}, \quad
\begin{bmatrix}
    0       \\
    0       \\
    1       \\
    \vdots  \\
    0       \\
    0
\end{bmatrix}, 
\quad
\cdots
\quad ,
\begin{bmatrix}
    0       \\
    0       \\
    0       \\
    \vdots  \\
    0       \\
    1
\end{bmatrix}
\]  
\end{center}
\caption{Left to right: one-hot vectors representing the first, second, third, and last opcodes in the instruction list, respectively.}
\label{fig:onehot}
\end{figure}

Numerous tasks with sequential inputs and/or sequential outputs can be modeled with RNNs~\cite{karpathy2015unreasonable}. For our application in smart contract opcode security threat detection, where inputs consist of a sequence of opcodes, opcodes are typically fed into the network in consecutive time steps. The most straightforward way to represent opcodes is to use a binary vector with length equal to the size of machine instruction list for each opcode in the directory---\emph{one-hot encoding}, as shown in Figure~\ref{fig:onehot}.

Such a simple encoding~\cite{Elman90findingstructure} has many disadvantages. First, it is an inefficient way of representing opcodes, as large sparse vectors are created when the number of instructions increases. On top of that, one-hot vectors do not capture any measure of functional similarity between opcodes in the encoding. Hence, we model opcodes with code vectors. It represents a significant leap forward in advancing the ability to analyze relationships between individual opcodes and opcode sequences. Code vectors are able to capture potential relationships in sequences, such as syntactic structure, semantic meaning, and contextual closeness. The LSTM learns these relationships when given a collection of supervised smart contract opcode data to initialize the vectors using an embedding algorithm~\cite{Tomas2013}. 

\begin{figure}[htbp]
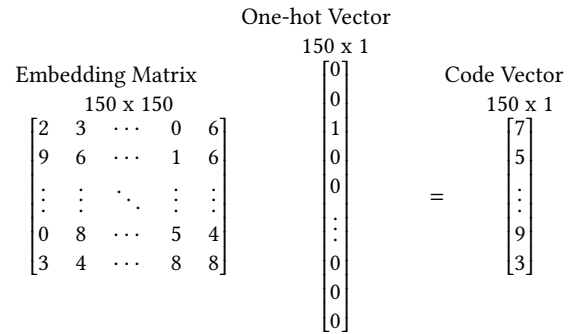

\begin{center}
    \[
\stackrel{\mbox{\parbox{3cm}{Embedding Matrix\\ \centerline{150 x 150}	}}}{%
\begin{bmatrix}
    2	&3	&\cdots	&0   &6	   \\
    9	&6	&\cdots &1   &6	   \\
    \vdots	&\vdots	&\ddots &\vdots   &\vdots	\\
    0	&8	&\cdots	&5   &4	   \\
    3	&4	&\cdots	&8   &8	   \\
\end{bmatrix}
}
\stackrel{\mbox{\parbox{2.5cm}{One-hot Vector\\ \centerline{150 x 1}	}}}{%
\begin{bmatrix}
    0       \\
    0       \\
    1       \\
    0       \\
    0       \\
    \vdots  \\
    0       \\
    0       \\
    0
\end{bmatrix}
}
=
\stackrel{\mbox{\parbox{2cm}{Code Vector\\ \centerline{150 x 1}	}}}{%
\begin{bmatrix}
    7       \\
    5       \\
    \vdots  \\
    9       \\
    3       
\end{bmatrix}
}
\]  
\end{center}
\caption{Example of Opcode Embedding.}
\label{fig:embedding}
\end{figure}

The embedding, shown in Figure~\ref{fig:embedding}, is a dense matrix in a linear space, which achieves two important functions. Firstly, by using an embedding with a much smaller dimension than the directory, it reduces the dimension of opcode representations from the size of the directory to the embedding size ($|\mathscr{U}| \ll |\mathscr{D}|$), where $|\mathscr{U}|$ and $|\mathscr{D}|$ are the embedding and directory sizes respectively. Secondly, learning the code embedding helps in finding the best possible representations, and groups similar opcodes in a linear space. 

A special case of the logistic function with output values from 0 to 1, sigmoid function, is used for the output layer. Intuitively, the outputs correspond to the probability that each opcode sequence is categorized as either one of the predicted class.

\section{Implementation}
Next, we turn to a discussion of how we implemented our LSTM detection tool in practice. We start by introducing the data source we used for vulnerable and not-vulnerable smart contracts. We then analyze the features and explain how we processed the contracts. Last we give details of how we trained the LSTM machine learning model.  

We trained and tested the proposed LSTM model on 620,000 contracts, by obtaining 920,179 existing smart contracts from the Google BigQuery~\cite{GoogleBigQuery} Ethereum blockchain dataset. This dataset includes the first block of Ethereum, up until block 4,799,998, which was the last block mined on December 26, 2017. 


\subsection{Data Source}
We used the Ethereum dataset downloaded from Google BigQuery. We then parsed the smart contracts' bytecode into opcode using the EVM instruction list~\cite{wood2014ethereum}. 

\subsubsection{Safe or Vulnerable}
In order to obtain labels for smart contracts in blocks 0 to 4,799,998, we ran the contracts through the \maian\ tool. A total of 920,179 contracts were processed, producing a number of flagged contracts. Processing our dataset using the \maian\ tool, we collected the sequential opcodes, which are instructions found in the EVM list of execution code, as inputs for our LSTM learning model. We then removed the wrongly flagged prodigal and suicidal contracts (false positives) identified by Nikolic et al.~\cite{Nikolic:2018:FGP:3274694.3274743}. The brief overview of the experiments to identify the false positives that were performed by the team of Nikolic et al. are as follows:
\begin{itemize}
    \item Concrete validation for prodigal and suicidal contracts was performed by running the flagged contracts along with its sequence of invocations produced by \maian\ on a private fork of Ethereum, effectively ensuring the reproducibility of the vulnerabilities. Contracts that were not exploitable but were flagged by \maian\ were categorized as false positives. 
    \item For contracts categorized as greedy (recall from section~\ref{greedy}), concrete validation was performed in a similar procedure for category (a) of greedy contracts by sending Ether to the flagged contracts and ensuring that no instructions exist within the contract that allowed the ether to be transferred out. For category (b), where instructions exist that allowed the possibility of Ether being transferred out, manual analysis was performed on the contracts which have source code available---none of them were identified to be true positives. 
\end{itemize}

Since no data were available for the wrongly flagged greedy contracts, we assumed all contracts in category (b) of the greedy contracts as false positives and removed them from our dataset, in accordance with findings presented in Nikolic et al.~\cite{Nikolic:2018:FGP:3274694.3274743}. After cleaning and processing our data, we report the number of distinct contracts, calculated by comparing the contract opcodes. Given the large number of flagged contracts, we proceeded to check for duplicates. We found 8640 distinct contracts that were flagged as suicidal (1207), prodigal (1461), greedy (5801), and both suicidal and prodigal (171). 

\begin{table}[htbp]
\renewcommand{\arraystretch}{1.5}
\resizebox{.95\linewidth}{!}{\begin{tabular}{|c|c|c|c|}
\hline
\textbf{Category}   & \textbf{\begin{tabular}[c]{@{}c@{}}Reported by~\cite{Nikolic:2018:FGP:3274694.3274743}\\ \maian\ Paper\end{tabular}} & \textbf{\begin{tabular}[c]{@{}c@{}}Processed by us\\ \maian\ Tool\end{tabular}} & \multirow{2}{*}{\textbf{\begin{tabular}[c]{@{}c@{}}Distinct\\ (flagged)\end{tabular}}} \\ \cline{1-3}
Contracts Processed & 970,898                                                                        & 920,179                                                                           &                                                                                        \\ \hline
Suicidal            & 1495                                                                           & 1544                                                                              & 1378                                                                                   \\ \hline
Prodigal (Leak)     & 1504                                                                           & 1786                                                                              & 1632                                                                                   \\ \hline
Greedy (Lock)       & 31,201                                                                         & 17,084                                                                            & 5801                                                                                   \\ \hline
\end{tabular}}
\caption{Processed and categorized contracts by \maian.}
\label{table:MAIAN_processed}
\end{table}

Table~\ref{table:MAIAN_processed} is a summary of data processed using the \maian\ tool. The difference of 50,719 processed contracts between the 970,898 contracts previously reported~\cite{Nikolic:2018:FGP:3274694.3274743} and the 920,179 processed by us was due to empty contracts. In addition, we believe that version updates of the \maian\ tool since the numbers were last reported in March 2018 contributed to this difference.

Using this dataset, we trained and tested the LSTM learning model on 8640 flagged contracts and 416,944 unflagged contracts, from which we removed invalid opcode instructions and duplicates. While \maian\ classifies vulnerable smart contracts into three categories of exploits, we considered these vulnerabilities as one class---vulnerable. In this two-class setting, each contract that is labeled as "0" in the \textit{category} field, Figure~\ref{fig:data_head}, indicates that it is not vulnerable. Otherwise, a vulnerable contract is labeled as "1".

We chose this security exploit detection task with this specific subset of the entire Ethereum blockchain dataset because of the public availability of smart contracts data, which has been symbolically analyzed~\cite{Nikolic:2018:FGP:3274694.3274743}, and it serves as a baseline for our model. 

\subsubsection{Opcode Features} 

As stated above in Section~\ref{sec:smartcontracts}, an Ethereum smart contract is a series of low-level EVM code that resides in the Ethereum blockchain. The EVM code, also known as bytecode, is a hexadecimal representation of a contract, which is something only the EVM can understand. Hence, we use a high-level language---Solidity to write smart contracts effectively. In order to deploy a smart contract, we compile the solidity code using a compiler, and it will translate our source code into bytecode. We then convert the bytecode into opcode, a human-readable format that is similar to any natural language. 

In the appendix of the Ethereum yellow paper~\cite{wood2014ethereum}, it contains a complete list of the EVM bytecode and its corresponding opcode. A bytecode to opcode disassembler~\footnote{\url{https://etherscan.io/opcode-tool}} can be used on any smart contract on the Ethereum blockchain to obtain the opcode. A fixed directory of 150 execution instructions for the smart contracts opcode is defined in the Ethereum yellow paper. Since 150 is a relatively small number when compared with most language sequence tasks in machine learning, all unique instructions were included for learning.

The EVM opcode is a machine language instruction that specifies the operations to be performed, and it reflects the logic of each smart contract. Opcodes have been successfully used in previous work to analyze various underlying issues of smart contracts~\cite{Chen:2018:DPS:3178876.3186046, 10.1007/978-3-030-15032-7_46}. Therefore, we expect that learning from a sequence of features extracted from opcodes is capable of detecting latent smart contract vulnerabilities. 

\subsubsection{Structural Properties}

\begin{figure}[htbp]
\begin{center}
\makebox[\linewidth]{\includegraphics[width=.95\linewidth]{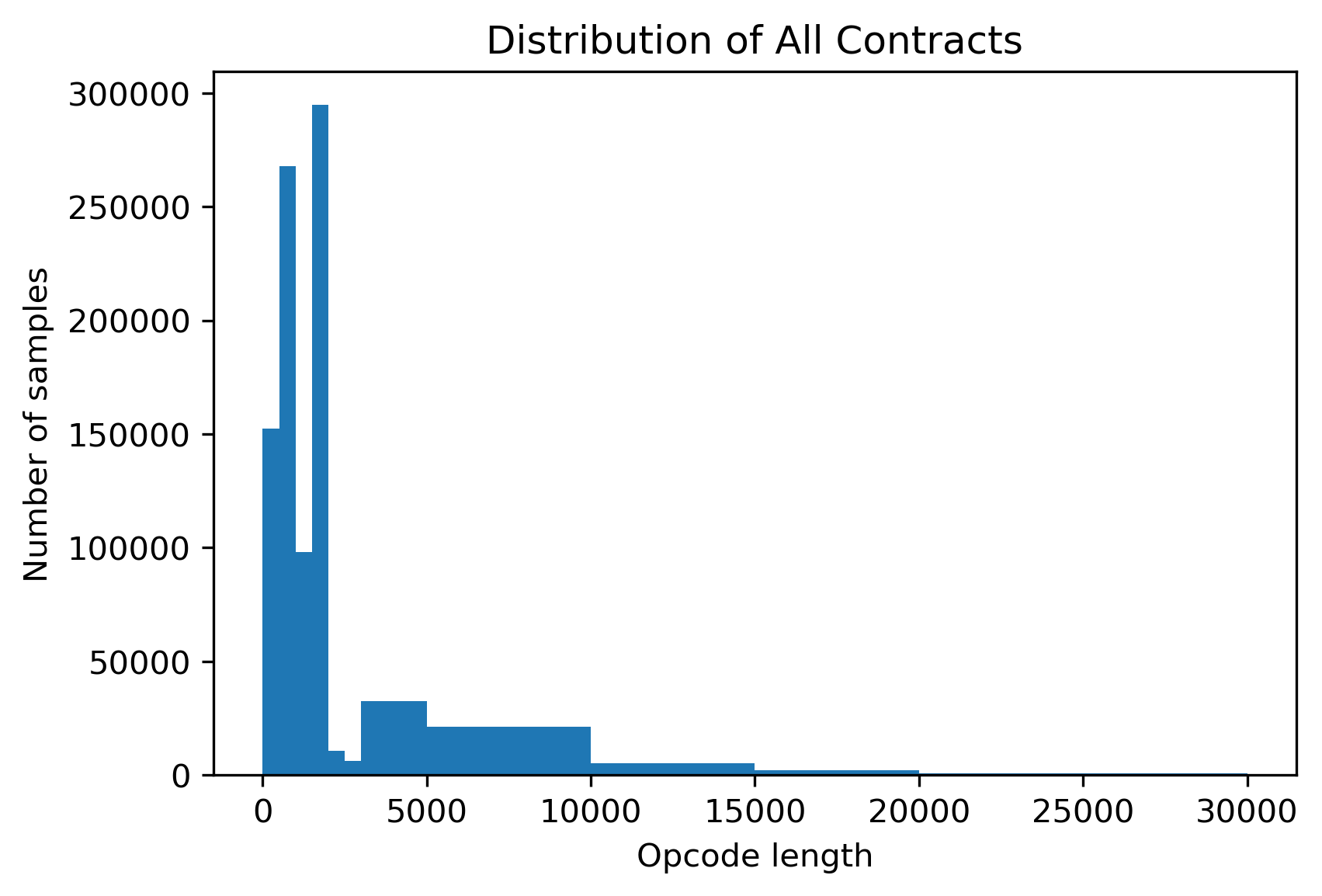}}
\end{center}
\caption{Histogram of the length of smart contract opcodes from original dataset. \textmd{Most of the smart contracts contain less than 1500 opcodes.}}
\label{fig:opcodehist}
\end{figure}

Figure~\ref{fig:opcodehist} shows some interesting characteristics of the length of the smart contracts we consider. Specifically, the histogram of the length of opcode each contract contains ranges from $2-53,936$. From the original dataset of approximately $900,000$ smart contracts, the opcode length statistics has a mode, median, and mean of $1541.0$, $1040.0$, and $1479.3$ respectively. For contracts that are not vulnerable, the statistics are very close to the population statistics, as roughly $99\%$ of the original dataset is made up of these contracts. 

On the other hand, the statistics of vulnerable contracts are significantly different than the not-vulnerable ones. Vulnerable contracts are reported to have a mode of $434.0$, a median of $818.0$, and a mean of $2648.6$. Hence, we decided to set the maximum length of LSTM opcode input to 1600. It is a design choice that would sufficiently cover most smart contracts, as 1600 is higher than the mode length of the entire population. Moreover, most vulnerable contracts would be detected since 1600 is much larger than both the mode and median length of the set of vulnerable contracts.  

\subsection{Data Processing}
While we collected a moderately large training set, it was highly imbalanced. It is an issue with classification problems where the classes are not represented equally, and one class outnumbers the other classes by a large proportion. Based on the distribution of the original dataset, $99.03\%$ of the contracts are labeled as not-vulnerable by \maian, while only $0.97\%$ of contracts are either greedy, suicidal, and/or prodigal. In order to handle the imbalanced set, we grouped all vulnerable contracts together to retrieve 8640 samples in one class, and samples not classified in any of the vulnerable categories were grouped into another class. Hence, samples are labeled as one of two classes, vulnerable or not-vulnerable.

Next, we resampled the dataset to achieve a balanced distribution, where half of the contracts are from the not-vulnerable class and the other half from the vulnerable class. We randomly sampled contracts from the not-vulnerable class set and created an equal number of synthetic vulnerable samples. Using a popular method to oversample minority classes, the Synthetic Minority Oversampling Technique (SMOTE)~\cite{Chawla02smote:synthetic}, we oversampled the minority (vulnerable) class and undersampled the majority (not-vulnerable) class. After performing the resampling, we ended up with a balanced dataset consisting of $50\%$ vulnerable and $50\%$ not-vulnerable contracts. 

In order to create synthetic samples, we first train a representative embedding using the original dataset. The embedding is a dense matrix that is a learned representation of the different opcodes. A smart contract sequence of opcode is then converted into to one-hot vectors. Using the learned embedding, we then perform the matrix dot product operation on the one-hot vectors, Figure~\ref{fig:embedding}. The resulting contract is now in dense code vectors. Finally, SMOTE oversampling is implemented on the code vectors of the vulnerable contracts, in order to generate synthetic samples.

\subsection{Training Details}

We trained our model with two LSTM layers of 128 and 64 hidden units respectively to learn from smart contracts with opcode length of 1600, and overcome both the vanishing gradient and long-term dependency issues, Section~\ref{sec:lstm_background}. The layers consist of a 150-dimensional word embedding with an input vocabulary of 150 opcode instructions. We found that our model was fairly easy to train on the balanced dataset. The classification task is based on a binary output using the sigmoid activation function. The LSTM generalizes well over our rebalanced training dataset, and it does not overfit the training samples. The resulting LSTM has 184,258 parameters, with training details as follows:
\begin{itemize}
	\item We divided the vulnerable class dataset of 8640 unique smart contracts into $64\%$ training, $16\%$ validation, and $20\%$ test. 
	\item We oversample the vulnerable training smart contracts into 200,000 samples.
	\item We then undersample an equal number of unique not-vulnerable contracts and add them to the training set.
	\item We use a total number of 620,000 smart contracts.
	\item We have a balanced training dataset of size 400,000, a validation set of size 100,000 (98,672 not-vulnerable, 1328 vulnerable), and a test set of size 120,000 (118,272 not-vulnerable, 1728 vulnerable).
	\item We use Adam~\cite{kingma:adam} as the adaptive gradient descent optimizer, and trained our LSTM model for a total of 256 epochs.
    \item We use batches of 256 smart contracts for the stochastic gradient descent optimizer to achieve speedy convergence.
    \item We use binary cross-entropy loss (log loss), which measures the performance of the classification model with output of a soft value between 0 and 1.
    \item We set the maximum input length to 1600 and zero-pad the contracts that were shorter than that.
\end{itemize}

\subsection{Evaluation Results}
In this section, we illustrate the experimental performance and results of our LSTM learning model on smart contract security threat detection tasks. Source code is available here~\footnote{\url{https://github.com/wesleyjtann/Safe-SmartContracts}}.

\subsubsection{Test Performance} \label{sec:balanced_set}
We use our LSTM learning model for evaluation and report the accuracy, recall, precision, $F_1$, and area under the curve of Receiver Operating Characteristic (AUC ROC) scores on the test dataset. The confusion matrix $C$, Figure~\ref{fig:confusion}, is used to evaluate classifier output quality. In binary classification, $C_{0,0}$ is the count of true positives, $C_{0,1}$ false positives, $C_{1,0}$ false negatives, and $C_{1,1}$ true negatives. 

\begin{figure}[tbp]
\begin{center}
\begin{tikzpicture}[
box/.style={draw,rectangle,minimum size=1.7cm,text width=1.3cm,align=center}]
\matrix (conmat) [row sep=.1cm,column sep=.1cm] {
\node (tpos) [box,
    label=left:\( \textbf{Vul.} \),
    label=above:\( \textbf{Vul.} \),
    ] {$C_{0,0}$ \\ 1602};
&
\node (fneg) [box,
    label=above:\textbf{Not-vul.},
    label=right:\(  \)] {$C_{0,1}$ \\ 340};
\\
\node (fpos) [box,
    label=left:\( \textbf{Not-vul.} \),
    ] {$C_{1,0}$ \\ 180};
&
\node (tneg) [box,
    ] {$C_{1,1}$ \\ 117,878};
\\
};
\node [left=.05cm of conmat,text width=1.3cm,align=right] {\textbf{Predicted \\ Class}};
\node [above=.05cm of conmat] {\textbf{Actual Class}};
\end{tikzpicture}
\end{center}
\caption{Confusion matrix of LSTM prediction.}
\label{fig:confusion}
\end{figure}

\begin{figure}[htbp]
\begin{center}
\makebox[\linewidth]{\includegraphics[width=.9\linewidth]{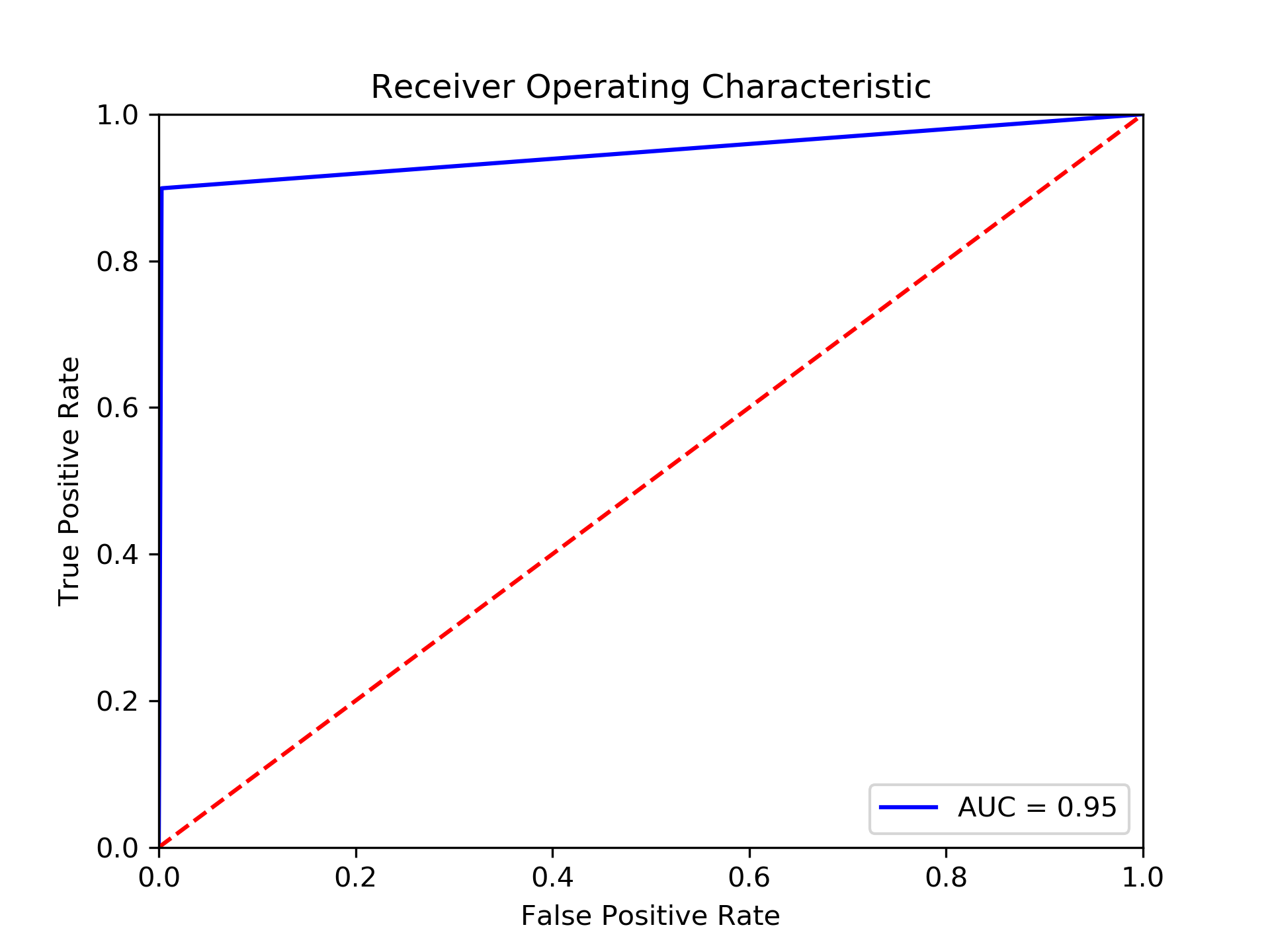}}
\end{center}
\caption{ROC Curve.}
\label{fig:ROC_550K}
\end{figure}

\begin{figure*}[htbp]
\centering
\begin{minipage}[b]{.48\textwidth}
\subfloat[][Accuracy]{\includegraphics[width=.9\linewidth]{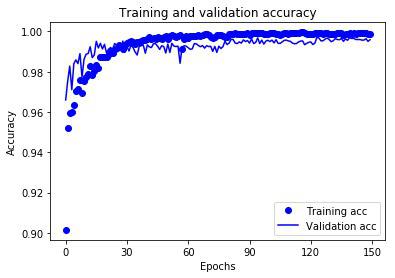}}
\label{fig:bal_accr}
\end{minipage}\qquad
\begin{minipage}[b]{.48\textwidth}
\subfloat[][Loss]{\includegraphics[width=.9\linewidth]{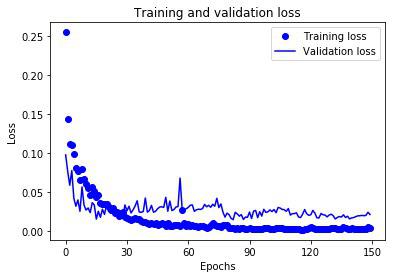}}
\label{fig:bal_loss}
\end{minipage}
\caption{Training and validation over 150 epochs.}
\label{fig:baltrain_and_val}
\end{figure*}

In addition, we present another measure that evaluates classifier output quality. The ROC curve, Figure~\ref{fig:ROC_550K}, features the true positive rate on the Y-axis and false positive rate on the X-axis. ROC curves are usually used to study the performance of binary classifiers. A large area under the curve (AUC) means a good classification performance. The training and validation process plots are shown in Figure~\ref{fig:baltrain_and_val}. We report classification accuracy and loss of the LSTM model on both the training and validation datasets.

From the results, we can see that the trained LSTM model achieved an impressive detection performance. For the test set, which consists of 120,000 samples, the LSTM model achieved a test accuracy of $99.57\%$. The test accuracy, Eq~\eqref{eq:accr_score}, is a straightforward detection evaluation metric. 

\begin{equation}\label{eq:accr_score}
    \text{Accuracy} = \frac{\text{True Positives} + \text{True Negatives}}{\text{Total Tested}}
\end{equation}

However, accuracy is not an adequate performance measure for highly imbalanced datasets, as models could easily achieve high accuracy by labeling all samples as the majority class. Since our test dataset mostly contains not-vulnerable contracts, our model is likely to neglect the minority (vulnerable) class. We proceed further to check both the precision and recall scores. The precision metric measures the ability of our LSTM model not to mislabel a safe smart contract as vulnerable, while the recall metric measures the model's ability to find all unsafe contracts.
We also present the $F_1$ score, Eq~\eqref{eq:f1}, that takes both precision and recall into account. It is the harmonic mean of the precision and recall scores, and it can also be interpreted as a weighted average of precision and recall, where an $F_1$ score reaches its best value at $100\%$ and worst score at $0\%$.

\begin{equation}\label{eq:f1}
    F_1 = 2  * \frac{\text{Precision} \times \text{Recall}}{\text{Precision} + \text{Recall}}
\end{equation}

Based on the LSTM performance, Table~\ref{table:lstm_perf}, it clearly indicates that our model is able to detect unsafe smart contracts accurately. On top of the detection accuracy, which was the only performance metric shown in~\cite{Nikolic:2018:FGP:3274694.3274743}, we provide additional evaluation metrics and present compelling evidence that the LSTM model is able to accurately detect unsafe smart contracts. 

Although the LSTM does not model \maian\ perfectly, we argue that this does not diminish the model's usefulness since it is likely due to the misclassification of the labels generated by the \maian\ tool. A sample of 69 unique contracts classified by the LSTM as vulnerable, which \maian\ analyzed as not-vulnerable, was checked against another symbolic analysis tool, \textsc{Securify}~\cite{Tsankov:2018:SPS:3243734.3243780}. We found that multiple contracts were categorized as vulnerable. Out of this small sample, 9 contracts were flagged as 'violation' and 2 contracts as 'warning'. We conjecture that the LSTM was able to generalize the content and exploits of smart contracts and detect vulnerabilities that \maian\ missed. In summary,

\begin{itemize}
	\item the LSTM model proved to be highly accurate at detecting unsafe smart contracts, and
	\item the LSTM with $99.57\%$ accuracy significantly outperformed \maian's $89\%$ accuracy in detecting unsafe smart contracts.
\end{itemize}

\begin{table}[htbp]
\resizebox{.9\linewidth}{!}{\begin{tabular}{|c|c|}
\hline
\textbf{Classification Performance Measure} & \textbf{\begin{tabular}[c]{@{}c@{}}LSTM \\ \%\end{tabular}} \\ \hline
Test Accuracy         & 99.57              \\ \hline
Recall Score          & 89.90              \\ \hline
Precision Score       & 82.49              \\ \hline
$F_1$ Score           & 86.04              \\ \hline
ROC AUC Score         & 94.81              \\ \hline
\end{tabular}}
\caption{LSTM detection performance measures.}
\label{table:lstm_perf}
\end{table}

We note that while symbolic analyzers are able to perform in-depth analysis of smart contract properties to detect the exact bugs, the LSTM model is able to quickly and accurately detect security exploits in the smart contracts. 
Moreover, concrete validation of flagged contracts in the case of \maian\ requires the creation of a private fork of the original Ethereum blockchain, and it can only analyze a contract from a particular block height where the contract is still alive. Not only is it a complicated and time-consuming task, but it is also not immediately clear how one selects a block height to include all flagged contracts that are still alive. Given that the LSTM model is a simple machine learning technique to implement, we recommend that it be used as the first line of defense, and complementing it with the concrete validation of symbolic analyzers if needed.


\subsubsection{FP Detection} \label{sec:addedFPs}
In addition, when we tested our learning model on the wrongly flagged contracts (false positives) identified in \cite{Nikolic:2018:FGP:3274694.3274743}, the LSTM is able to detect many of them. By using the corrected labels of these wrongly flagged samples in our training, it helps our model improve its ability to accurately learn from challenging and complex smart contracts. 

\begin{table}[htbp]
\resizebox{.95\linewidth}{!}{\begin{tabular}{|c|c|c|c|}
\hline
\textbf{Category} & \textbf{\begin{tabular}[c]{@{}c@{}}\# Distinct contracts\\ tested\end{tabular}} & \textbf{\begin{tabular}[c]{@{}c@{}}\# Correctly\\ classified\end{tabular}} & \textbf{\begin{tabular}[c]{@{}c@{}}Accuracy \\ \% \end{tabular}} \\ \hline
Suicidal FPs        & 14   &   13   & 92.86   \\ \hline
Prodigal FPs        & 35   &   28   & 80.00    \\ \hline
Greedy FPs          & 41   &   31  & 75.61     \\ \hline
\end{tabular}}
\caption{LSTM classification performance on contracts falsely flagged by \maian\ tool.}
\label{table:lstm_FP}
\end{table}

We first corrected the labels of the 27,266 FP contracts and processed them to check for duplicates by comparing the opcode between smart contracts. Out of these 27,266 contracts, 451 of them were found to be unique. As duplicates that simultaneously appear in training, validation, and test sets could lead to disingenuous performance results, we only use unique smart contracts in our datasets. Next, we divided these 451 FP contracts into the same proportion of $64\%$ (288), $16\%$ (73), and $20\%$ (90), and added them to the training, validation, and test sets respectively. With no overlap of unique contracts in each dataset, the FP contracts in the training set are then resampled to roughly $5\%$ of the entire training set. The results in Table~\ref{table:lstm_FP} demonstrates that the LSTM model is able to detect a large number of these complicated smart contracts.

\subsubsection{Analysis Runtime} \label{sec:runtime}

\begin{figure}[htbp]
\begin{center}
\makebox[\linewidth]{\includegraphics[width=.95\linewidth]{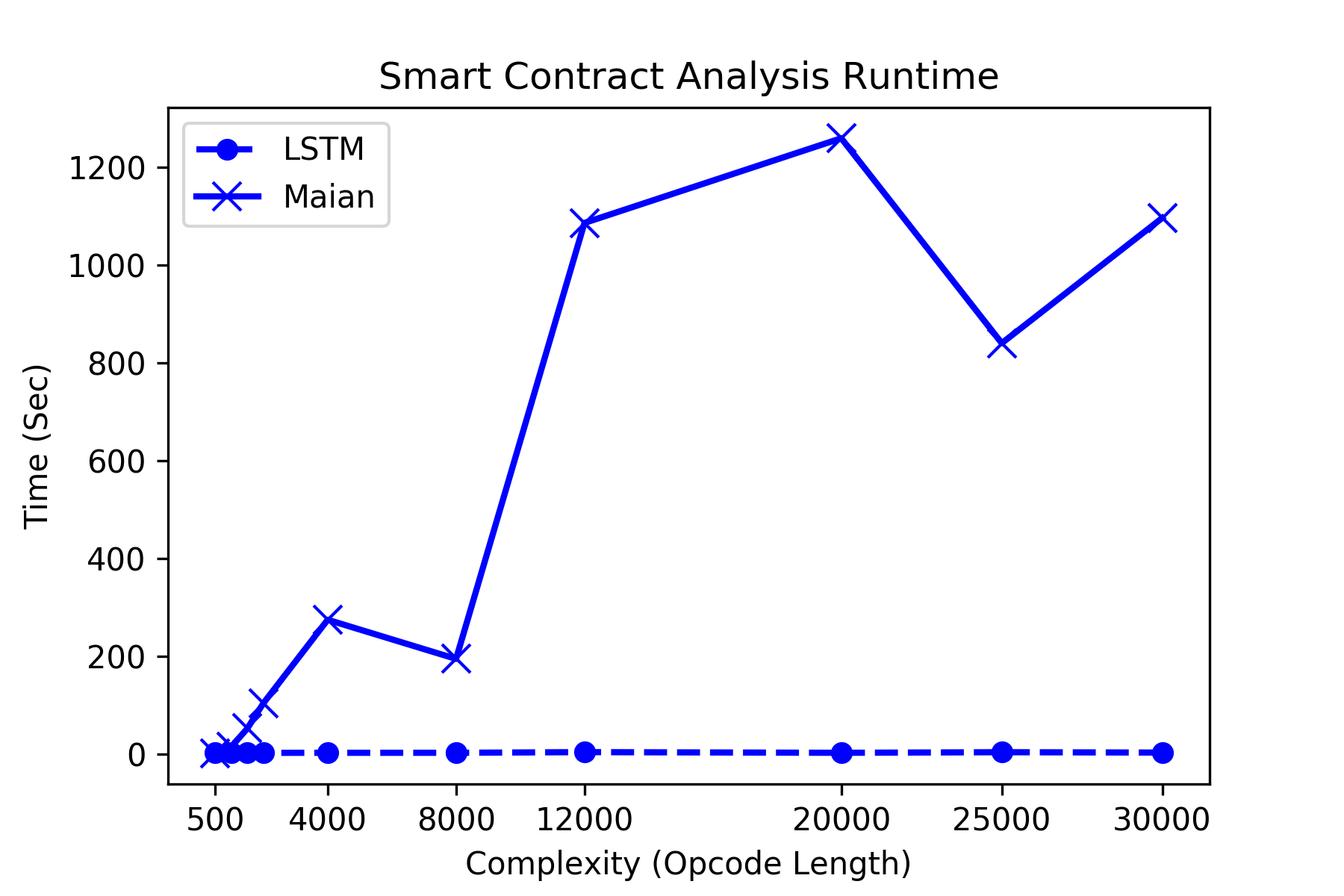}}
\end{center}
\caption{Analysis time of LSTM vs \maian. \textmd{}}
\label{fig:runtime}
\end{figure}

Figure~\ref{fig:runtime} demonstrates that the analysis time of symbolic tool \maian\ increases rapidly with the complexity of the smart contracts, measured by opcode length. On the other hand, the time required by the LSTM remains fairly constant. To evaluate the different analysis times taken by the LSTM and \maian, we sample 5 smart contracts each from 9 different opcode lengths: [500, $1000$, $1500$, $2000$, $4000$, $8000$, $12,000$, $20,000$, $25,000$]. We then measure the average analysis time of the 5 contracts of each length and plot the graph to highlight the difference between our model and \maian. 

In our study, we ran the experiments with Intel Xeon E5-2698 v4 2.2GHz CPU, 512GB system memory, and NVIDIA Tesla V100 GPU. It took the LSTM around an average of 2.2 seconds to analyze contracts of opcode length 30,000, while \maian\ required 1096.2 seconds. The results provide strong evidence that our model can easily scale to a desirable level, in order to process a massive amount of smart contracts growing every day. 

\section{LIMITATIONS}
The limits of the LSTM machine learning classifier can be attributed to the following challenges in detecting if a smart contract contains security exploits.  

Our LSTM model assumes the sequence of opcode features, which reflects the logic of a smart contract, can generalize its content and its vulnerabilities. This assumption is sometimes violated since learning of opcode sequences is unable to consider the control-flow (e.g., loops, internal function calls, etc.) of a smart contract. It does not support the checking of data- and control-flow properties. 
For example, integer overflows (or underflows) which may occur due to iterative loops within a contract or computation of balance via function calls can affect the correctness of the contract. In some cases, these control-flow vulnerabilities can be exploited to siphon funds from a contract in an unjust manner~\cite{banisadr_2018}.
In such situations, the LSTM will fail to perform well against vulnerable smart contracts since control-flow analysis is outside its domain of sequence learning. 

The categories of vulnerabilities defined in~\cite{Nikolic:2018:FGP:3274694.3274743} may also be identified as a limitation on what the LSTM model can achieve. It has been observed that numerous characterization of exploits in smart contracts is not limited to the three classes, Section~\ref{sec:vulnerablecontracts}.
Apart from the 3 classes of vulnerabilities, Luu et al.~\cite{Luu:2016:MSC:2976749.2978309} studied 4 other classes of vulnerabilities---transaction-ordering dependence, timestamp dependence, mishandled exceptions, and re-entrancy. Those classes were further expanded upon by Zhou et al.~\cite{8328743}. New risks such as transaction origin risk and zero division risk were added. As for Kalra et al.~\cite{DBLP:conf/ndss/KalraGDS18}, they focused on a different set of vulnerabilities involving the semantic correctness and higher-level business logic fairness properties of contracts. 
As observed, while there are other classes of vulnerabilities defined, it is generally not a reasonable assumption for one system to capture all exploits at any point in time. Exploits and attacks evolve. The impact of this behavior on the LSTM is that contracts which contain similar exploits to those categories defined in \maian, will be classified as vulnerable, when they actually may not since the exploits are previously unseen by the LSTM model. While this does lower the detection performance of the machine learning model, we argue that this does not reduce its usefulness since these misclassifications represent contracts that are still at considerable risk and require attention.  

Last, the binary classification design also presents another main limitation. Although it allows for the fast and light detection of smart contract vulnerabilities, the LSTM does not provide additional insights for analysis. There may be a need for further analysis after a vulnerable contract is identified. It would be desirable to determine the part of contract code that is diagnosed as vulnerable in order to allow for supplementary manual analysis, however no attempt to do this have been made in our experiments; but it deserves further work.

\section{RELATED WORK}
There has been an increasing adoption of deep learning in the field of computer security. More specifically, a variety of security attack detection scenarios have benefited from the advancement of deep learning techniques. Du et al.~\cite{Du:2017:DAD:3133956.3134015} proposed the modeling of a system log as a natural language sequence using a deep LSTM model, in order to perform the task of anomaly detection in systems. A system event log is designed to record system states at critical points, and it is produced by a program that follows a rigorous set of logic and control flows. Since system logs contain text that is largely unstructured and distinct, analytics is challenging. However, the LSTM anomaly detection model was able to achieve remarkable performance by learning and encoding entire log messages (including the timestamp, log key, and parameter values).  

In a similar fashion, Shen et al.~\cite{Shen:2018:TPS:3243734.3243811} showcased the importance of using the sequence memory architecture in recurrent neural networks for the task of predicting security events in a computer. With the increased complexity of malicious activities in computer systems, simple methods such as the Markov Chains~\cite{norris_1998} or 3-gram models~\cite{Brown:1992:CNG:176313.176316} are no longer effective in predicting these malicious events. By leveraging on the long-term memory typical of LSTM models, it has been shown that their system was able to predict future events of a machine, specific steps that an attacker would undertake, based on previous observations. 

Shin et al.~\cite{shin190918} investigated function identification, a crucial step in many binary code analysis techniques, for malware detection and fixing vulnerable software. They showed that recurrent neural network architectures such as the LSTM can identify functions in binaries with great accuracy and efficiency. 
Going a step further, Chua et al.~\cite{chua203650} 
trained a recurrent neural network to learn function type signatures from disassembled binary code without prior knowledge of the compiler or instruction set. It is the first machine learning based system that targets function signature recovery while maintaining "comparability" of its learned outcomes to conventions used by other analysis tools.

However, to the best of our knowledge, the detection of smart contract vulnerabilities based on sequence learning has not been investigated so far. The tools used in smart contract analysis are mainly symbolic ones based on formal methods of verification~\cite{Luu:2016:MSC:2976749.2978309, DBLP:conf/ndss/KalraGDS18, Tsankov:2018:SPS:3243734.3243780, Liu:2018:RFR:3183440.3183495, Nikolic:2018:FGP:3274694.3274743}.

\section{Conclusion}
We have presented a scientific work---based on machine learning towards safer smart contracts---that is able to effectively comb through a large number of contracts for security threats. Our results suggest that sequential learning of smart contract security threats provides significant improvements over symbolic analysis tools, achieving detection accuracy of $99.57\%$ and $F_1$ score of $86.04\%$. Furthermore, up to $92.86\%$ of particularly challenging contracts that were otherwise deemed as false positives by \maian, were correctly detected by our sequential learning model. Our machine learning approach is the first of its kind that maintains its analysis time as smart contracts grow in complexity. This makes it possible to build a scalable smart contract security threat detection tool based on machine learning.

In future work, we plan to study the impact of state-of-the-art sequence modeling techniques on smart contract threat detection. The Transformer---a new simple network architecture that dispenses with recurrence entirely and is based solely on attention mechanisms---has been shown to be superior to recurrent neural networks~\cite{NIPS2017_7181}. Further experiments using attention models might allow smart contract security exploits to be detected more effectively. While we measure detection performance based on classification correctness with respect to given labels, a remaining challenge is obtaining more accurate labels to improve detection proficiency of models during training.

\begin{acks}
We thank Aashish Kolluri and Prof. Prateek Saxena for sharing the validation data, and all our anonymous reviewers for their valuable comments and helpful suggestions. 
\end{acks}

\bibliographystyle{ACM-Reference-Format}
\bibliography{biblography}



\end{document}